\documentclass[article,onecolumn,superscriptaddress,nofootinbib]{revtex4}
\usepackage{xcolor,color}
\usepackage[a4paper,
            bindingoffset=0.2in,
            left=.7in,
            right=1in,
            top=1in,
            bottom=1in,
            footskip=.25in]{geometry}
\usepackage[utf8]{inputenc}
\usepackage{float}
\usepackage[T1]{fontenc}
\usepackage{latexsym,amssymb,amsmath,amsfonts}
\usepackage{graphicx}
\usepackage{mathrsfs}
\usepackage{float}
\usepackage{xcolor}
\usepackage{hyperref}
\usepackage{graphicx}
\usepackage{subcaption}
\hypersetup{colorlinks=true, linkcolor=magenta, citecolor=green }
\usepackage{enumerate}

\begin{document}

\title{Anisotropic quantum polytropes}

\author{Hermano Velten}%
\email{hermano.velten@ufop.edu.br}
\affiliation{%
Departamento de F\'isica, Universidade Federal de Ouro Preto (UFOP), Campus Universit\'ario Morro do Cruzeiro, 35.400-000, Ouro Preto, Brazil}%

\author{Felipe S. Escórcio}
\email{escorcio@ufop.edu.br}
\affiliation{%
Departamento de F\'isica, Universidade Federal de Ouro Preto (UFOP), Campus Universit\'ario Morro do Cruzeiro, 35.400-000, Ouro Preto, Brazil}%

\author{Nadson J. S. Trindade}
\email{nadysonsilvaa@gmail.com}%
\affiliation{Instituto de Física, Universidade Federal da Bahia, 40210-340 Salvador, BA, Brasil
}%
\date{\today}

\begin{abstract}The structure of astrophysical objects is usually modeled under the assumption of hydrostatic equilibrium. However, actual configurations may deviate from perfect spherical or isotropic properties. Consequently, cosmic objects are expected to exhibit some degree of anisotropy. This consideration also extends to hypothetical dark structures, such as dark stars and dark matter halos. Although the nature of dark matter remains unknown, axion-like particles (ALPs) are strong candidates, suggesting that dark matter halos may have originated from bosonic configurations undergoing gravitational collapse, sustained by boson-boson interactions in the condensate state. This system is described by the Gross-Pitaevskii-Poisson equation. Furthermore, within the framework of the Bohm-de Broglie approach, quantum effects—encapsulated in the so-called quantum potential—may play a significant role in equilibrium astrophysical configurations. In this study, we examine a class of static anisotropic boson stars which are non-minimally coupled to gravity. By including all these factors, we derive a generalized Lane-Emden-like equation and conduct a detailed analysis of the maximum degree of anisotropy that such systems can sustain, thereby identifying physically viable equilibrium configurations. Apart from focusing on the impact of anisotropic contributions, we find that for the so-called Quantum Polytropes (when the quantum potential is the main responsible for the equilibrium condition), the anisotropic factor and the gravitational field have opposite roles compared to the classical case. This leads to a new class of hydrostatic equilibrium objects.
\end{abstract}

\maketitle

\section{Introduction}
The interplay between the gravitational interaction and the quantum regime is a matter of intense debate and investigation in the literature. One can look at both the emergence of gravitational effects at the quantum scale as well as explore possible manifestations of quantum properties at macroscopical physical systems subject to gravity. The ideal formulation would be a comprehensive framework that seamlessly integrates the principles of general relativity and quantum theory. This unified theory is anticipated to offer a thorough description of spacetime's microstructure at the Planck scale, where the fundamental constants $c$ (the speed of light in a vacuum), $\hslash$ (the reduced Planck's constant), and $G$ (Newton's gravitational constant) converge to define natural units of mass, length, and time. Since this desired stage has not yet been reached, we are left with phenomenological attempts to incorporate both realms of physics.

In one of the possible realizations of this program, the Schrödinger equation is naturally applied in providing a detailed description of the temporal evolution of quantum systems within Newtonian gravitational potentials. Prominent examples include atomic interferometry experiments \cite{APeters_2001, PhysRevLett.125.191101, PhysRevLett.98.111102}, recent studies examining Bose-Einstein condensates under microgravity conditions \cite{PhysRevLett.110.093602}, and investigations into quantum superposition at macroscopic scales \cite{kovachy2015quantum}. Beyond these experimental approaches, broader frameworks addressing quantum gravitational effects have been investigated in astrophysical and cosmological settings. Phenomena such as galaxy rotation curves and the accelerated expansion of the universe, traditionally interpreted using theories like Newtonian mechanics and general relativity, often necessitate the introduction of unknown concepts like dark matter and dark energy to reconcile theory with observations. Therefore, incorporating quantum principles into these scenarios can be seen as an interesting way to offer novel perspectives.

By considering that the structure of astrophysical systems can be connected somehow to quantum aspects, one usually adopts the Gross-Pitaevskii-Poisson system and its variants \cite{chavanisI.84.043531,chavanisII, haegeman2017quantum,guzman2006gravitational,Velten:2011ab}. Recent works (see Refs. \cite{Schwabe:2016rze,Calabrese:2016hmp,hui2017ultralight,Bar:2018acw,korshynska2023dynamical,chavanis2017dissipative,Niemeyer:2019aqm,PhysRevD.100.083022,Hui:2021tkt,Ferreira:2020fam} and the references therein for a concise collection of articles on the topic) associate dark matter halos with ultralight particles, whose properties can be described by solitonic solutions, such as those derived from Bose-Einstein condensates. With a Bohmian perspective, such condensates can explain the density distribution at the center of galactic halos and address discrepancies in the standard cold dark matter (CDM) model \cite{Boehmer:2007um}.

De Broglie-Bohm mechanics also stands out by employing a hydrodynamic formulation for quantum systems, enabling the extension of quantum mechanics to different scales and scenarios. This formulation allows for the description of self-gravitating systems at scales that, under appropriate conditions, can achieve hydrostatic equilibrium. Classic examples of the latter include stars, modeled as spherically symmetric objects where radial pressure counterbalances gravity. The structure of galaxies (and their dark halos) and galaxy clusters can also be designed with this description. Such systems, described by the Lane-Emden equation, are characterized as polytropic objects with specific equations of state \cite{abellan2020double, Herrera:1997plx}. The introduction of quantum features into these astrophysical systems has led to the development of isotropic quantum polytropic models \cite{Heyl:2017jei}, enabling, for example, the description of dark matter halos as quantum scalar fields and offering stable solutions for the density profile in galaxies, as well as explaining phenomena such as the formation of dense galactic cores \cite{del2009density}. Therefore, the extended Schrödinger-Poisson system studied in this work expands the possibilities for the modeling of dark matter halos and structure formation, and can represent the Newtonian limit from a more fundamental relativistic field theory.

Despite the favorable possibility described above, observational evidence indicates that many astrophysical systems, such as rotating galaxies, exhibit significant anisotropies. Galactic rotation and properties such as dynamic viscosity affect the mass distribution and, consequently, the hydrostatic equilibrium. For example, in neutron stars, phenomena like rotation, interior magnetic fields, and exotic matter or condensates are also included as sources of anisotropies. To address these situations, it is essential to incorporate a degree of anisotropy in modelling astrophysical spherically symmetric configurations in hydrostatic equilibrium \cite{Bowers:1974tgi,Isayev:2017rci,Dev2004, herrera2020newtonian,Leon:2023yvz,Kumar:2022kho,Andrade:2023jvm,Becerra:2024wku}. As a result, it is also shown in the literature that anisotropy can affect the general properties of relativistic stars \cite{Mak2003,Dev2004,Raposo2019}. A recent sequence of articles with a complete description of self-gravitating anisotropic fluids in both Newtonian and relativistic gravity is found in Refs. \cite{Cadogan:2024mcl,Cadogan:2024ohj,Cadogan:2024ywc}.

The investigation we propose here is also motivated by the fact that the quantum potential is equivalent to anisotropic pressure in spherically symmetric hydrostatic equilibrium configurations as pointed out by Refs. \cite{Boehmer:2007um,chavanisI.84.043531}. This suggests that classical anisotropic configurations and isotropic quantum-supported systems can give rise to degenerate states.

For this purpose, the present work initially introduces a formulation for isotropic quantum polytropes based on a Gross-Pitaevskii-Poisson system, later extending its formalism to construct anisotropic quantum polytropes. Subsequently, the application of the Gross-Pitaevskii system is highlighted. In this context, subsequent sections also discuss the application of the Gross-Pitaevskii-Poisson system, with extensions to include relevant anisotropies, thereby exploring the dynamics and stability of astrophysical systems that can be represented by the hydrodynamical equilibrium analysis.

\section{Towards anisotropic self-gravitating quantum polytropes}

\subsection{Gross-Pitaevskii-Poisson system}
If we consider a bosonic system in the limit where the temperature $ T \to 0$, all bosons in the system condense into a single quantum state, giving rise to the Bose-Einstein Condensate (BEC). In this regime, the behavior of the many-body system can be adequately described by a collective wave function $\Psi(\mathbf{r}, t)$, which satisfies the following dynamic equation, known as the Gross-Pitaevskii equation:
\begin{eqnarray} \label{GP}
i\hslash\dfrac{\partial\Psi(r,t)}{\partial t} = -\dfrac{\hslash^2}{2m}\nabla^2\Psi(r,t) + V(r,t)\Psi(r,t) + \frac{4\pi \hslash^2 a_s}{m}\vert\Psi(r,t)\vert^2\Psi(r,t),
\end{eqnarray}
which is a nonlinear version of the Schrödinger equation for a particle with mass $m$ subject to interactions determined by the s-wave boson–boson scattering length $a_s$. With astrophysical applications in mind, an N-particle self-gravitating system should be subject to the Poisson equation.
\begin{equation}\label{Poisson}
\nabla^2 V(r,t)= 4 \pi G \rho = 4 \pi G N m |\Psi|^2,  
\end{equation}
and, hence, equations \eqref{GP} and \eqref{Poisson} are known as Gross-Pitaevskii-Poisson system. Moving to the fluid perspective, one applies the polar form of the Madelung transformation $\Psi(r,t) = A(r,t)e^{iS(r,t)}$. Within this scenario, $\rho$ is associated with the squared modulus of the wave function and is interpreted as the probability density. The fluid motion, with velocity $\mathbf{u} = \nabla S / m$, is considered irrotational, i.e., $ \nabla \times \mathbf{u} = 0$, as long as the phase $S$ is single-valued and differentiable.

The Gross-Pitaevskii-Poisson system is separated into its real and imaginary components, matching a hydrodynamic formulation of the Madelung equations
\begin{equation}
\frac{\partial \rho}{\partial t}+\nabla \cdot (\rho {\bf u})=0,
\end{equation}
\begin{equation}\label{eqS}
\frac{\partial S}{\partial t}+\frac{(\nabla S )^2}{2m}+V(r,t) +Q_{eff}=0,
\end{equation}
with all quantum corrections encoded in the effective quantum potential
\begin{equation}\label{Qeff}
Q_{eff} = Q_q+Q_{s}=-\dfrac{\hbar^2}{2m}\frac{\nabla^2 \sqrt{\rho}}{\sqrt{\rho}} + \frac{4\pi \hslash^2 a_s}{m}\rho.
\end{equation}
 This definition of effective quantum potential ($Q_{eff}$) takes into account the usual quantum potential 
 \begin{equation}
     Q_q=-\dfrac{\hbar^2}{2m}\frac{\nabla^2 \sqrt{\rho}}{\sqrt{\rho}}=-\frac{\hbar^2}{4m} \left[ \frac{\nabla^2 \rho}{\rho} - \frac{1}{2} \frac{(\nabla \rho)^2}{\rho^2} \right],
 \end{equation} 
 represented by the first term in the right-hand-side of definition \eqref{Qeff} as well as a potential $Q_{s}$ associated with the boson-boson interaction i.e., $Q_{s}\propto a_s$. Note that for $(a_s > 0)$, both contributions could eventually cancel each other out, but this is a finely tuned situation.

Next, taking the gradient of \eqref{eqS} and using the irrotational flow condition one finds 
\begin{equation}\label{Euler}
\frac{\partial {\bf u}}{\partial t}+({\bf u}\cdot\nabla){\bf u} =-\nabla V(r,t) -\frac{1}{m}\nabla Q_{eff},
\end{equation}
which resembles the Euler equation. At this point, a common assumption in the literature is to associate the gradient of the boson-boson term in \eqref{Qeff} to an effective pressure. This approach leads to the interpretation BEC configurations can be mimicked by polytropic fluids with equation of state
\begin{equation}
P_{s}=\frac{2\pi\hslash^2 a_s}{m^3}\rho^2.
\end{equation}

In many astrophysical and fluid dynamics applications, one particularly useful approach is the polytropic equation of state, which is commonly used to model the behavior of gases and self-gravitating systems \cite{Ngubelanga:2015ith,PolytropicConstancia,Godani:2023wrq}. This equation is given by  
\begin{equation}\label{EoS}
P = K \rho^{\gamma}, \quad \gamma = 1 + \frac{1}{n},    
\end{equation}
where $K$ is a proportionality constant known as the polytropic constant. The parameter $\gamma$ is the adiabatic index, which depends on the polytropic index $n$ as above.  

For a specific case where the polytropic index takes the value $n = 1$, we obtain $\gamma = 2.
$
In this scenario, the polytropic constant is given explicitly by  $
K_{s} = \frac{2 a_s \hbar^2}{m^3},$  
where $a_s$ is a system-dependent coefficient, $\hbar$ is the reduced Planck's constant, and $m$ represents the mass of the constituent particles. This pressure can be linked to an effective potential, which can be expressed as
\begin{equation}\label{gamma2}
    Q_{s}(\rho) = K_{s}\frac{\gamma }{\gamma - 1} \rho^{\gamma - 1} \quad \overset{\gamma=2}{\rightarrow}\quad     Q_{s}(\rho) = 2 K_{s} \rho. 
\end{equation}

This expression plays a crucial role in the analysis of equilibrium configurations, stability, and dynamical evolution of astrophysical structures, including stars, gaseous disks, and even certain cosmological models \cite{tooper1964general, toomre1964gravitational}. By modifying the polytropic index $n$, one can explore a wide range of physical scenarios, from isothermal to adiabatic conditions, making this formulation highly versatile in theoretical and computational studies.  

An interesting remark already presented in \cite{Boehmer:2007um,chavanisI.84.043531} is the identification of a relationship between the quantum potential and the quantum stress (or pressure) tensor. This tensor formalism provides a framework for understanding the quantum forces that act on a system
\begin{equation}
    -\frac{1}{m} \nabla Q_q \equiv -\frac{1}{\rho} \partial_j P_{ij},
\end{equation} 
where $P_{ij}$ is the quantum stress (or pressure) tensor. This equation shows that the force exerted by the gradient of the quantum potential can be equivalently expressed as a term involving the divergence of the quantum stress tensor. Note that this is only valid in the absence of the gravitational field $V$. This establishes a link between the scalar potential and the anisotropic pressure described by the tensor.
The definition of the quantum stress tensor $P_{ij} $ can be recast in the form
\begin{equation}
    P_{ij} = -\frac{\hbar^2}{4m^2} \rho \partial_i \partial_j \ln \rho=\frac{\hbar^2}{4m^2} \left( \frac{1}{\rho} \partial_i \rho \partial_j \rho - \delta_{ij} \Delta \rho \right)
\end{equation}

While previous works have established a correspondence between the quantum potential and an effective anisotropic pressure, our approach departs from this identification. In the present analysis, we do not assume that the pressure is necessarily equal to the quantum potential. This generalization allows for a broader class of hydrostatic equilibrium configurations and enables a more flexible modeling of quantum effects beyond the standard interpretation based on the quantum stress tensor alone.

\subsection{``Quantum" astrophysical configurations in hydrostatic equilibrium}

In order to model astrophysical systems we start rewriting the Euler equation such that
\begin{equation}
\label{Euler2}
\frac{\partial {\bf u}}{\partial t}+({\bf u}\cdot\nabla){\bf u} =-\frac{\nabla P}{\rho} -\nabla V(r,t) -\frac{1}{m}\nabla Q_{eff}.
\end{equation}
By writing the above equation differently from \eqref{Euler}, we are including an extra {\it ad hoc} dynamical ingredient, the ``macroscopic" pressure $P$. The latter can be connected to the typical physical aspects of the stellar interior e.g., thermal pressure, radiation pressure and degeneracy pressure, while $P_{s}$ refers to an extra effective source of pressure related to the boson-boson interaction ($\propto Q_{s}$). With this interpretation we are adopting a different approach in comparison to the one in Ref. \cite{chavanisI.84.043531}. We assume that apart from the contributions from quantum origin encoded in the effective potential $Q_{eff}$, the system is also subject to a pressure $P$ as any other astrophysical configuration. We therefore have more generality than the previous investigations in the literature.

Still, based on the approach presented in \cite{chavanisI.84.043531,chavanisII}, we analyze a gravitational system composed of bosons that remain condensed in a single quantum state at an appropriate scale, forming a self-gravitating structure that can exhibit either isotropy or anisotropy. Initially, we assume that the dynamics of the system are described by a wave function $\Psi(\mathbf{r}, t)$, whose evolution is governed by an equation that includes both the gravitational potential $V(\mathbf{r}, t)$ and a short-range interaction. Thus, the total potential of the system is the sum of the gravitational potential $V$ and an effective potential accounting for short-range interactions. 

For the static systems studied in this work, the time-independence of  $|\Psi|^2$ and  $\rho$ implies that the gravitational potential is time-independent too,  $V \equiv V(r)$. Hence, in the stationary case, the complete system is described by a set of equations known as the Gross-Pitaevskii-Poisson (GPP) system:
\begin{equation} \label{GGP}
-\nabla\cdot\left(\frac{\nabla P}{\rho}\right)+\nabla^{2}\Big(\dfrac{\hbar^2}{2m^2}\frac{\nabla^2 \sqrt{\rho}}{\sqrt{\rho}}\Big) - \frac{4\pi \hslash^2 a_s}{m^2}\nabla^2\rho= 4\pi G\rho.
\end{equation}

Here, the Poisson equation is coupled to the effective quantum potential, while the density $\rho(r,t)$ is associated with the matter density\footnote{The quantum equilibrium condition is a central postulate in the de Broglie-Bohm interpretation. It states that the initial distribution of the particles is given by $|\Psi(r,t=0)|^2$. Since the guidance equation preserves this distribution over time, the matter density $\rho(r,t)$ always coincides with
$|\Psi(r,t)|^2$.} of the polytrope. Explicitly, $\rho = |\psi|^2$, as defined from the wave function $\Psi$. The normalization of 
$\Psi$ does not directly affect the trajectory, since the guiding equation is invariant under multiplication by a global constant (as $\nabla\Psi/\Psi$ is independent of the norm). However, the particle density in the ensemble (the initial statistical distribution) is postulated as $|\Psi|^2$. 

When considering spherical symmetry, we have
\begin{equation}\label{GeneralnoAnis}
-\frac{1}{r^2} \frac{d }{d r}\left(\frac{r^2}{\rho}\frac{d P}{d r}\right)+\frac{\hbar^2}{2m^2}\dfrac{1}{r}\dfrac{d^{2}}{dr^{2}}\left[ \dfrac{1}{\sqrt{\rho}}\dfrac{d^{2}}{dr^{2}}(r\sqrt{\rho})\right] - \frac{4\pi \hslash^2 a_s}{m^2} \frac{1}{r^2}\dfrac{d}{dr}\left[r^2\frac{d\rho}{dr} \right] = 4\pi G\rho.
\end{equation}
This is the most general hydrostatic equilibrium condition for an isotropic sphere subject to quantum interactions given by $Q_q$ and $Q_s$ in which the internal pressure $P$ is bounded by the gravitational attraction. In the next subsection we discuss how to include anisotropic features into this configuration.

\subsection{Including anisotropic pressure}
In isotropic systems, the pressure is the same in all directions at a specific point. However, in fluids where anisotropy is present, the pressure can vary with direction, implying that the local hydrostatic equilibrium is altered. Our strategy, then, is to address the hydrostatic equilibrium by considering the  potential $Q_{eff}+V$, characteristic of the gravitational quantum polytrope, but with the additional anisotropy term $\Delta$ defined as the difference between the tangential and radial pressure components
\begin{equation}
    \Delta=P_t-P_r.
\end{equation}

Therefore, considering $d {\bf u}/dt = 0$ in \eqref{Euler2}, we have 
\begin{equation} \label{7.1}
\frac{d P}{dr} =  - \rho\frac{ d V}{dr}- \rho \frac{d  Q_{eff}}{dr} + \dfrac{2}{r}\Delta .
\end{equation}

This equation illustrates that the outward pressure gradient from the material sector is counterbalanced by an inward gravitational force as well as by the quantum contribution encoded in $Q_{eff}$ and an anisotropic force, whose impact varies with their signs. When tangential pressure surpasses radial pressure, the anisotropy is positive, producing an outward force that opposes gravitational attraction, leading to a repulsive effect on the astrophysical object. Conversely, a negative anisotropy factor results in an inward force, reinforcing gravitational pull.

Equation \eqref{7.1} can be regarded as the Newtonian approximation of the Tolman–Oppenheimer–Volkoff equation for anisotropic matter in General Relativity, or, equivalently
\begin{equation} \label{Full}
\nabla P = -\rho \nabla \Phi_{eff}^{ani}= - \rho \frac{d  Q_{eff}}{dr} - \rho\frac{ d V}{dr} + \dfrac{2}{r}\Delta .
\end{equation}
Inspired by the formulation presented in Ref. \cite{herrera2020newtonian}, the above equation can be recast in a more compact version such as 
\begin{equation}
\Phi_{\rm eff}^{anis}=Q_{eff}+V-\int^{x=r}_{x=0}\frac{2 \Delta}{x \rho}dx.   
\end{equation}
According to the above definition the anisotropic contribution can be understood as a potential-like term as well as the quantum potentials contributing to $Q_{eff}$ and the gravitational potential $V$.

A general model for anisotropic contribution is \cite{Herrera:1997plx,Bowers:1974tgi}
\begin{equation}\label{Delta}
\Delta \equiv (\rho + P_r) r^N f.  
\end{equation}

The anisotropy in the interior may exhibit spatial variations and, in many cases, depends nonlinearly on the radial pressure. These effects are incorporated through a function $ f\equiv f(P_r,r)$,which is taken as $f = 1$ in this analysis to simplify the algebraic treatment. However, this choice does not compromise the generality of the model, as the nonlinearity will be captured through the functional dependencies of the density ($\rho$) and pressure ($P$), whose specific forms will be determined by the properties of the investigated structure. The combination $(\rho +P)$ is important since one expects the anisotropy should vanish at the polytropic radii, exactly where the density and pressure also vanish. It is observed that, for ($N > 1$), the obtained solution satisfies the boundary condition previously established in \cite{Bowers:1974tgi}. As expected, this means the anisotropy increases with the radius. Moreover, the particular choice of ($N = 2$) is associated with an anisotropy that vanishes at the origin with a quadratic behavior, ensuring regularity at the center of the spherically symmetric configuration, also shown in Ref. \cite{Bowers:1974tgi}. The case $N=2$ also applies to slowly rotating systems. Finally, for the polytropic case, we will consider the usual form $P_r= P = K \rho^{1+1/n}$.

Adding the anisotropic term to the general case described in \eqref{GeneralnoAnis} and adopting 
\begin{equation}\label{btorho}
 b \equiv b(r)= \sqrt{\rho(r)},
\end{equation}
the general hydrostatic equilibrIum condition with quantum corrections and the anisotropic contribution reads
\begin{equation}
\frac{\Omega_P}{r^2}\frac{d}{dr}\left(r^2 b^{2\gamma-1}b^{\prime}\right)=  \dfrac{\Omega_q}{r}\dfrac{d^{2}}{dr^{2}}\left[ \dfrac{1}{b}\dfrac{d^{2}}{dr^{2}}(rb)\right] - \frac{\Omega_s}{r^2}\dfrac{d}{dr}\left[r^2\frac{d \ b^2}{dr} \right] -\Omega_G b^2 + \dfrac{\Omega_A}{r^{2}}\dfrac{d}{dr}\left(\dfrac{2r\Delta}{b^2}\right) .
\end{equation}

The magnitude of pressure contributions is encoded in the parameter $\Omega_{P}\equiv2K\gamma$. In the above equation, $\Omega_A$ is a dimensionless parameter to quantify the magnitude of the anisotropic contribution. Obviously, one recovers the isotropic case with $\Omega_A =0$.   
After expanding the derivatives and defining $\Omega_q = \frac{\hslash^2}{2m^2}$, $\Omega_s = \frac{4\pi \hslash^2 a_s}{m^2}$, and $\Omega_G = 4\pi G$, we note that the following relation applies: $\Omega_q=\frac{\Omega_s}{8\pi a_s}$. The dimensionless parameter $\Omega_P$ is introduced to facilitate the study of the influence of the pressure gradient $\nabla P$. Similarly, the free ad hoc dimensionless parameter $\Omega_A$ is included here. By setting $\Omega_A = 0$, the isotropic configuration is restored. By varying $\Omega_A$, one can examine the anisotropic effects represented by $\Delta$. Hence, this is equivalent to,
\begin{eqnarray}\label{general}
\Omega_P \delta_P(b)=\Omega_q\left(b''''+\frac{4b'''}{r} - \frac{2b'''b'}{b}
-\frac{8b''b'}{rb} -\frac{b''^2}{b}+ \frac{4b'^3}{rb^2} + \frac{2b''b'^2}{b^2}\right)\\ \nonumber-4\Omega_s\left(\frac{bb'}{r}+\frac{b'^2}{2}+\frac{bb''}{2}\right) -\Omega_G b^3+\Omega_A \delta_A(b)  ,
\end{eqnarray}
where the contribution of the pressure gradient (recalling the polytropic equation of state of the type $P = K \rho^{\gamma} = K b^{2\gamma}$) can be written according to
\begin{equation}\label{deltaP}   
    \delta_P(b) =\frac{2b^{2\gamma-3}b^{\prime}}{r}+\left(2\gamma-3\right)b^{2\gamma-4}b'^2+b^{2\gamma-3}b^{\prime\prime}.
\end{equation}
Also, the contribution originated from the anisotropic term is given by
\begin{equation}\label{deltaA}
    \delta_A(\Delta) = \frac{2b}{r^2}\left[\frac{\Delta}{b^2} + \frac{r\Delta'}{b^2}-\frac{2r\Delta b'}{b^3} \right],
\end{equation}
where, in terms of the new definitions provided above, we can rewrite $\Delta$ \eqref{Delta} and its first derivative as,
\begin{eqnarray}
\Delta&=&\left[1+\frac{\Omega_P}{2\gamma} b^{2(\gamma-1)}\right]b^2r^2, \\ \Delta^{\prime}&=&\frac{\Omega_P}{2}\left[(2\gamma-1)b^{2\gamma-1}r^2  b^{\prime}+2b^{2\gamma} r\right]+2bb^{\prime}r^2
+2b^2 r.
\end{eqnarray}

Equation \eqref{general}, incorporating the contributions described in \eqref{deltaP} and \eqref{deltaA}, represents the complete formulation of hydrostatic equilibrium, including relevant quantum effects such as the quantum (Bohmian) potential, boson-boson interaction, pressure sources, and anisotropic features, in addition to gravitational-like interactions described by the Poisson equation.
As expected, one can easily verify that the contribution from the boson-boson interaction (the term proportional to $\Omega_s$) is equivalent to that obtained from the polytropic equation of state pressure gradient when adopting $\gamma = 2$.

\section{Results for the polytropic structure}

In this section we present numerical analysis of all subcases of the full hydrostatic equilibrium equation \eqref{general}. We shall solve equation \eqref{general} numerically in order to obtain the function $b(r)$ which immediately provides the radial density profile via \eqref{btorho}.  

\subsection{Anisotropic classical polytropes}
\begin{figure}[t!]
    \centering
    \begin{minipage}[t]{0.48\columnwidth}
        \centering
    \includegraphics[width=\columnwidth]{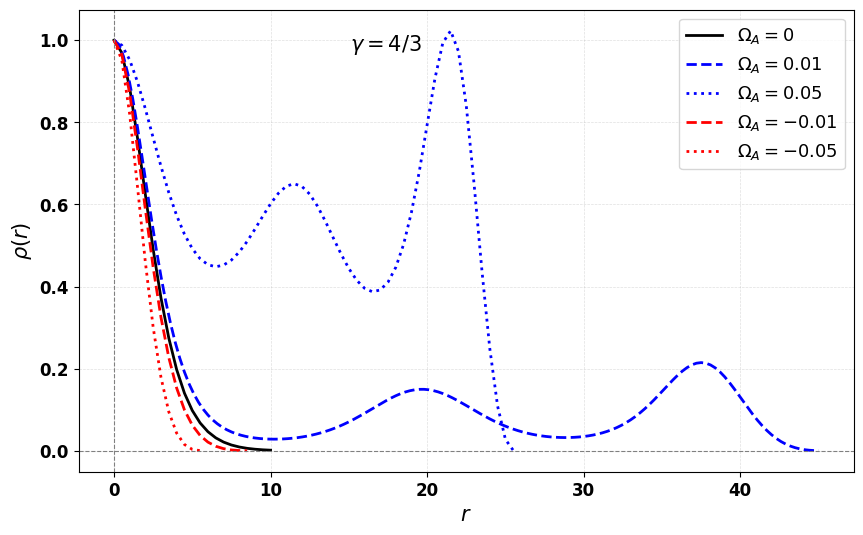}
    \end{minipage}%
    \hfill
    \begin{minipage}[t]{0.48\columnwidth}
        \centering
    \includegraphics[width=\columnwidth]{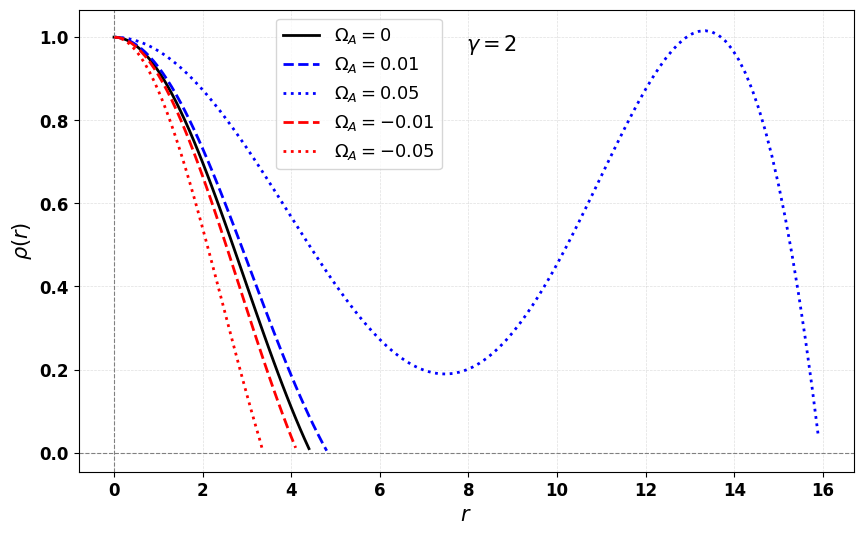}
    \end{minipage}
    \hfill
        \begin{minipage}[t]{0.48\columnwidth}
        \centering
    \includegraphics[width=\columnwidth]{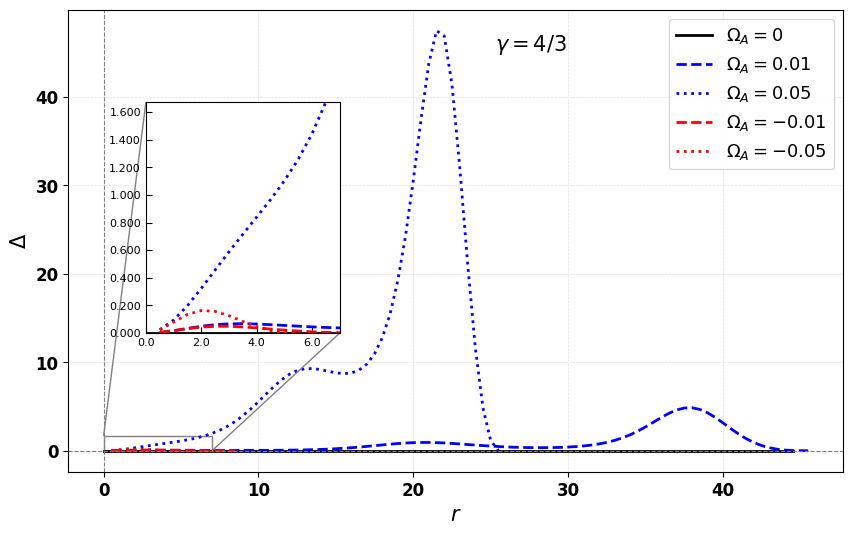}
    \end{minipage}%
    \hfill
    \begin{minipage}[t]{0.48\columnwidth}
        \centering
\includegraphics[width=\columnwidth]{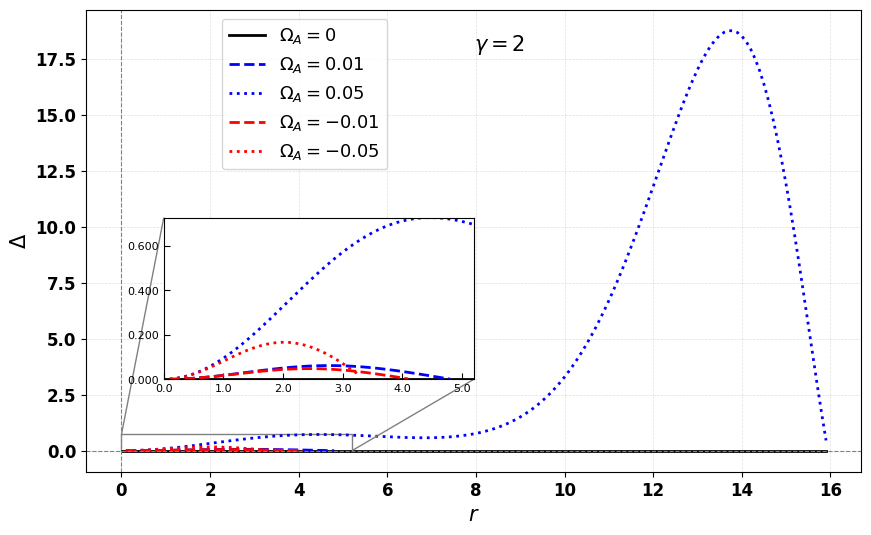}
    \end{minipage}
    \caption{Top panels: Radial density profile in the non-interacting case with $\Omega_q = 0$, $\Omega_G=1$ and $\Omega_p = 1$ adopting $\gamma=4/3$ (left panel) and $\gamma=2$ (right panel). Bottom panels: Anisotropic contribution $\Delta$ profile in the non-interacting case with  $\Omega_q = 0$.}
    \label{fig:GAMMAclas}
\end{figure}

To establish a calibrated starting point, we first focus on standard polytropes without quantum effects influencing the hydrostatic equilibrium equation. This configuration is obtained by setting $\Omega_q=\Omega_s=0$ in \eqref{general} leading it to a second order differential equation. This is the classic scenario studied in astrophysics textbooks, where the outward pressure gradient is counterbalanced by the gravitational attraction.
We shall consider two polytropic indices: $\gamma=4/3$ ($n=3$) and $\gamma=2$ ($n=1$), as shown in Figure \ref{fig:GAMMAclas}. The density profile is normalized to units of the central density at the core of the configuration, i.e., $\rho(r=0)=\rho_c=1$. As usual, to avoid numerical instabilities, the boundary conditions are defined at the point $r_0 \rightarrow 0$. In our numerical computations, we set $r_0=0.001$, then $\rho(r_0)=1$ or, equivalently, $b(r_0)=1$
For the first derivative, we adopt $b^{\prime}(r_0)=0$, since this is expected to be a local maximum of the density function. The horizontal axis represents the radius in arbitrary units.

The polytropic index $n = 3$, the relativistic limit of a completely degenerate gas, is a well-justified choice in astrophysical modeling due to its applicability to both relativistic degenerate matter and stellar structure. In the case of massive white dwarfs, the equation of state for relativistic degenerate matter supports the use of $n = 3 $, as it accurately represents the pressure-density relation in their cores. Additionally, within main-sequence stars like the Sun, the same index is employed in the radiation zone, aligning with the Eddington standard model of stellar structure. This broad applicability demonstrates that $n = 3$ provides a physically meaningful representation of both compact and main-sequence stellar objects, making it a valuable tool in astrophysical analysis. On the other hand, a polytrope with index $n = 1$ is commonly used to model self-gravitating isothermal spheres, making it a valuable tool in astrophysics. This index is particularly relevant in cases where the pressure is directly proportional to density, such as in certain phases of star formation and the structure of low-mass, fully convective stars. Additionally, an $n = 1 $ polytrope provides useful approximations for systems where thermal equilibrium is maintained over long timescales. Its simplicity allows for analytical solutions. As stated before, this index also corresponds to the BEC polytrope as in \eqref{gamma2}.

We present the radial density profile $\rho(r)$ in both top panels of Figure \ref{fig:GAMMAclas}. The isotropic configurations ($\Omega_A=0$) are represented by the black line, while positive anisotropic configurations ($\Omega_A>0$) are shown in blue and negative ones ($\Omega_A<0$) in red. This color scheme will be maintained in the subsequent figures. Also, the adopted free parameter values appear in the figure's caption. 
A notable feature is the oscillatory behavior in the density profile induced by positive values of the anisotropic parameter, approximately of the order $\Omega_A \sim\mathcal{O}(10^{-2})$. We have checked that assuming $\gamma=4/3$ There are bound structures up to $\Omega_A\sim 2.0$. With $\gamma=2$ there are no bound structures for $\Omega_A \gtrsim 0.07$.

We have also replicated this analysis with $\Omega_P=10$ (meaning the pressure magnitude is increased by a factor of 10). As a result, the equilibrium configuration expands, reaching larger radii by a factor $\sim 3$.
In both cases, the anisotropic factor produces the same qualitative effects: negative values ($\Omega_A<0$) compress the polytropic configuration, whereas positive values ($\Omega_A>0$) tend to lead to unstable configurations, as expected. The bottom panels of Figure \ref{fig:GAMMAclas} display the radial profile of the quantity $\Delta$, which represents the anisotropic contribution. The direct relationship between $\Delta$ and the oscillatory features in the density profile is now evident.

\subsection{Anisotropic quantum polytropes}
Next, we start understanding how the quantum contributions influence the equilibrium configurations.  
By activating the quantum potential contribution, we will set $\Omega_q=1$, while still keeping $\Omega_P=1$ and adopting the same polytropic indices as before. By switching on $\Omega_q$, we now have to solve a fourth-order differential equation, and other initial conditions should be specified. We fix $b^{\prime \prime}(r_0) = -1$, since we have demanded the density has a local maximum at $r_0$. Also, as far as we have checked, $b^{\prime \prime \prime}(r_0)$ can assume any arbitrary value without changing the results.  

\begin{figure}[t!]
    \centering
    \begin{minipage}[t]{0.48\columnwidth}
        \centering
        \includegraphics[width=\columnwidth]{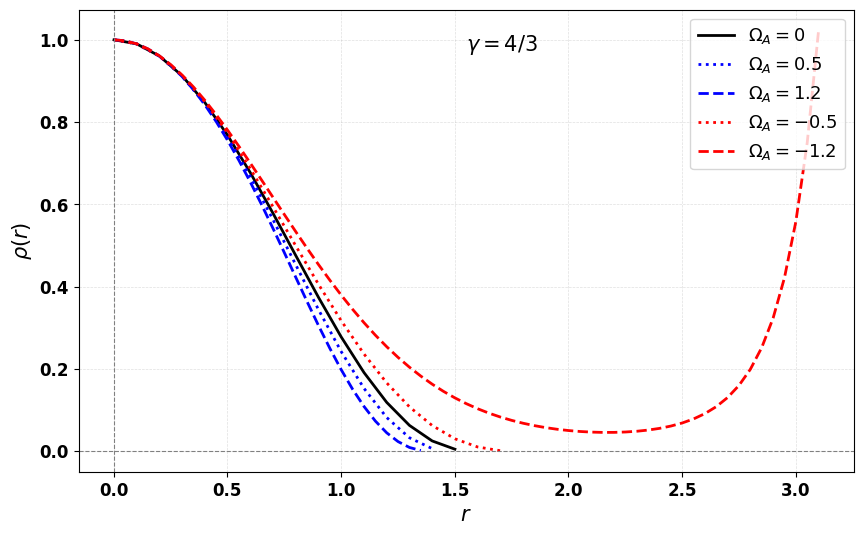}
    \end{minipage}%
    \hfill
    \begin{minipage}[t]{0.48\columnwidth}
        \centering
\includegraphics[width=\columnwidth]{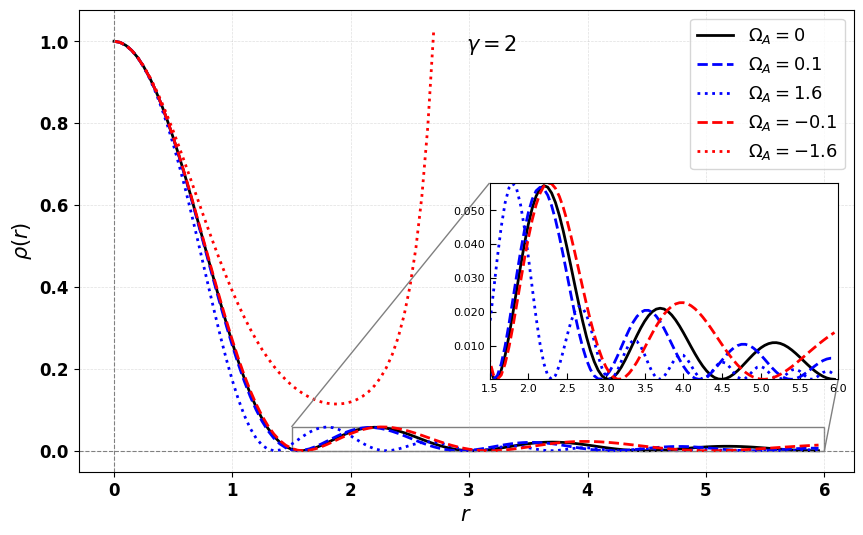}
    \end{minipage}
    \caption{Radial density profile in the non-interacting case with $\Omega_q = 1$, $\Omega_G=1$ and $\Omega_p = 1$ adopting $\gamma=4/3$ (left panel) and $\gamma=2$ (right panel). When $\gamma=4/3$, the limiting situation occurs when $\Omega_A \lesssim -1.0$. With $\gamma=2$, the function no longer reaches $\rho(r) = 0$ for values close to or less than $\Omega_A = -1.4$.}
    \label{fig:GAMMAQ}
\end{figure}

We show this analysis in Fig.~\ref{fig:GAMMAQ}. By comparing this figure with Fig.~\ref{fig:GAMMAclas}, we notice that the quantum potential is much more relevant for the polytropic structure. For both polytropic indices, the radius is compressed to the values $R \sim 1.5$. Also, when including the anisotropic contribution, we notice that the radial profile is less sensitive to the $\Omega_A$ factor, i.e., values $|\Omega_A|=0.5$ change the equilibrium radius by a small factor.  

\begin{figure}[ht!]
    \centering
    \begin{minipage}[t]{0.48\columnwidth}
        \centering
        \includegraphics[width=\columnwidth]{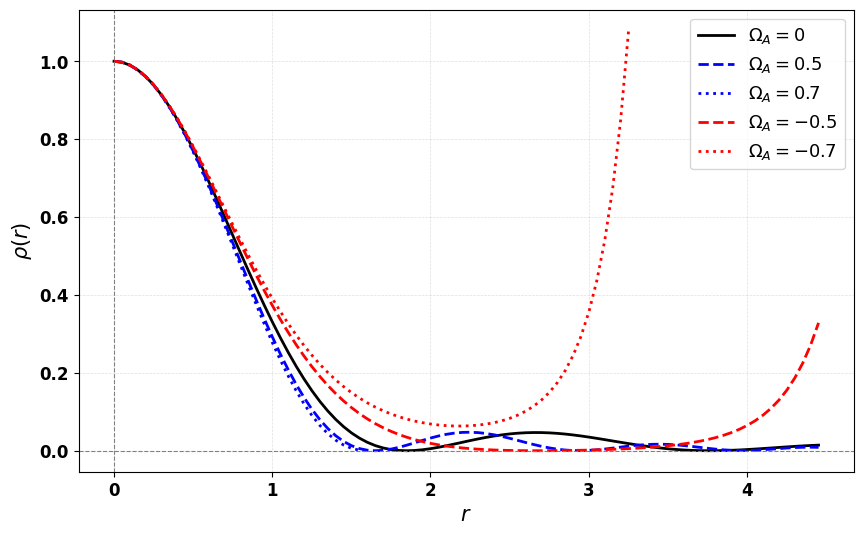}
        \subcaption{Non-interacting case ($\Omega_s=0$, $\Omega_q = 1$, $\Omega_p = 0$, $\Omega_G = 1$)}
        \label{fig:sub}
    \end{minipage}%
    \hfill
    \begin{minipage}[t]{0.48\columnwidth}
        \centering
        \includegraphics[width=\columnwidth]{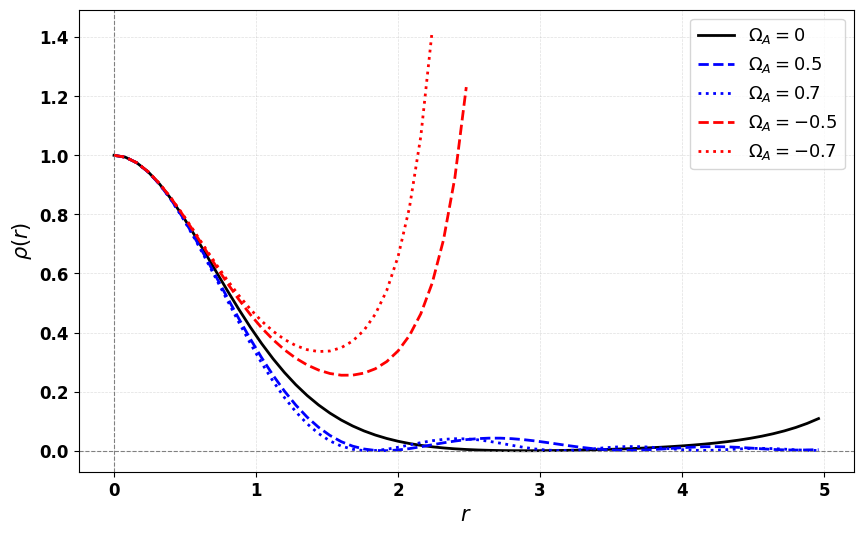}
        \subcaption{Non-interacting case ($\Omega_s=0$, $\Omega_q = 1$, $\Omega_p = 0$, $\Omega_G = 10$)}
        \label{fig:sub}
    \end{minipage}%
    \hfill
    \begin{minipage}[t]{0.48\columnwidth}
        \centering
        \includegraphics[width=\columnwidth]{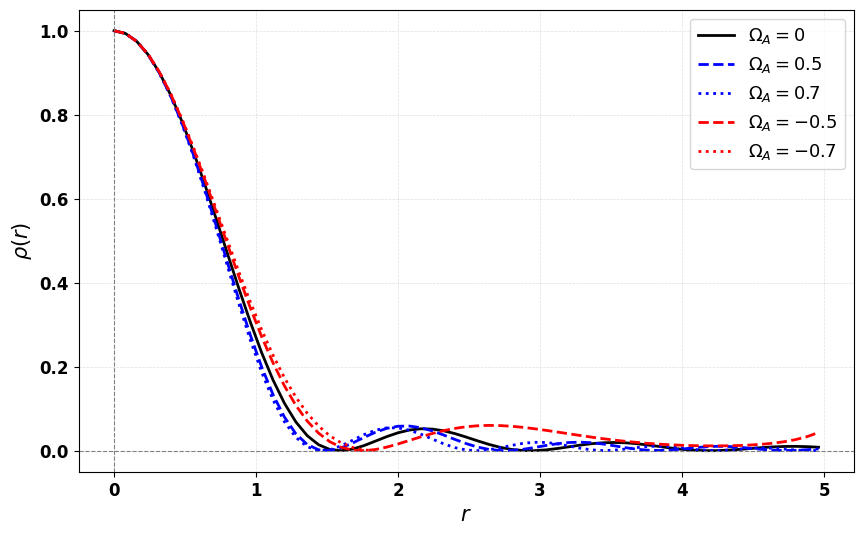}
        \subcaption{Non-interacting case ($\Omega_s=0$, $\Omega_q = 1$, $\Omega_p = 0$, $\Omega_G = -10$)}
    \end{minipage}
    \caption{Radial density profiles for different parameter combinations in the pressureless case ($\Omega_P = 0$).}\label{fig:G}
\end{figure}

Another noteworthy case is $\Omega_P = 0$, which corresponds to the anisotropic, pressureless Schrödinger-Poisson system. This system represents the anisotropic version of the model studied in \cite{Heyl:2017jei}, referred to as "Quantum Polytropes" in that reference. It is obtained from \eqref{general} by setting $\Omega_P = \Omega_s = 0$. Different configurations of this system are shown in Fig.~\ref{fig:G}.

We describe now the anomalous behavior of the quantum polytropes, which is the main aspect found in this work. We start noticing a different impact of the anisotropic factor. Now, positive $\Omega_A$ values shrink the polytropic radius, as seen in all panels of Fig.~\ref{fig:G}.  

The top-left panel in figure \ref{fig:G} presents the numerical results for this case. It is important to note that, in the absence of pressure gradient effects, the polytropic index does not play any role here. The gravitational collapse is counterbalanced solely by the repulsive interaction provided by the quantum potential. In this figure, the red dotted line represents the limiting configuration, indicating that bound configurations exist only for $\Omega_A \gtrsim -0.5$.

To reinforce the anomalous behavior of this system, in Fig.~\ref{fig:G} we arbitrarily change the magnitude of the gravitational interaction adopting $\Omega_G=+10$ (top right panel) and $\Omega_G=-10$ (bottom panel) in the pressureless non-interacting case. In the first case ($\Omega_G=+10$) the structure becomes less bounded, in opposition to what is expected if the magnitude of the gravitational interaction is increased. With $\Omega_G=-10$ (a repulsive interaction) the structure is compressed.

Concerning the influence of the magnitude of pressure effects, we have also checked that different $\Omega_P$ values have only a mild impact on the equilibrium configurations when fixing $\Omega_q=1$. But, by increasing the $\Omega_P$ values the polytropic radius decreases, in opposition to the case with $\Omega_q=0$.

\begin{figure}[h]
    \centering

    \begin{minipage}[t]{0.48\columnwidth}
        \centering
        \includegraphics[width=\columnwidth]{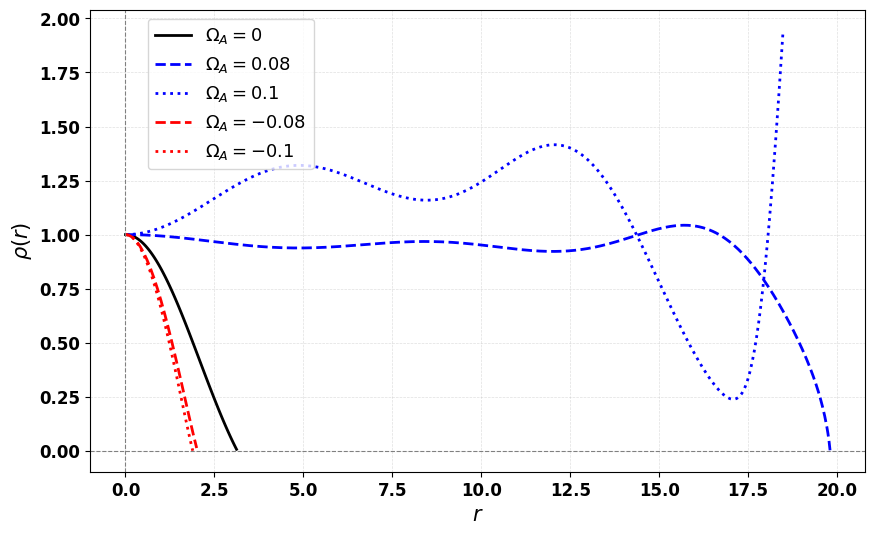}
        \subcaption{Interacting regime in the Thomas–Fermi limit ($\Omega_s=1$, $\Omega_q = 0$, $\Omega_p = 0$)}
        \label{fig:sub1}
    \end{minipage}%
    \hfill
    \begin{minipage}[t]{0.48\columnwidth}
        \centering
        \includegraphics[width=\columnwidth]{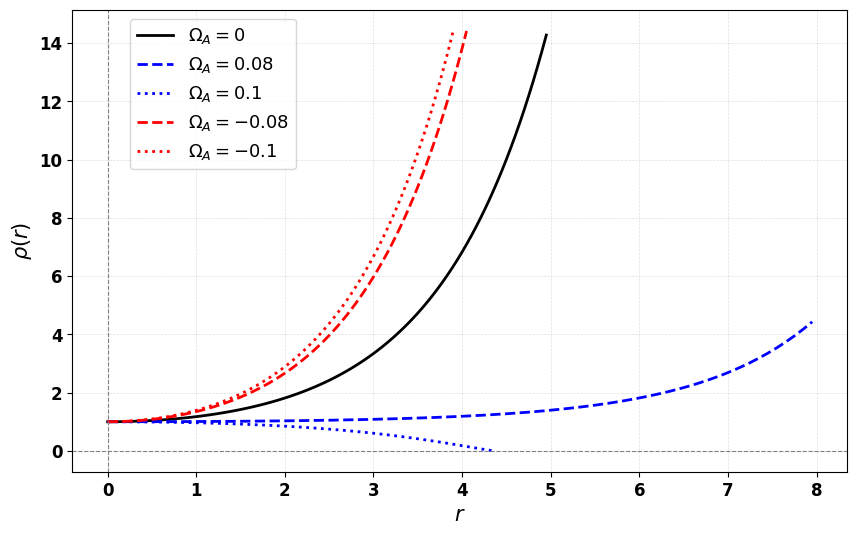}
        \subcaption{Interacting regime in the Thomas–Fermi limit ($\Omega_s=-1$, $\Omega_q = 0$, $\Omega_p = 0$). There are bound states (i.e., $\rho(R) = 0$) for $\Omega_A \gtrsim 0.09$}
        \label{fig:sub2}
    \end{minipage}

    \vspace{0.5cm}

    \begin{minipage}[t]{0.48\columnwidth}
        \centering
        \includegraphics[width=\columnwidth]{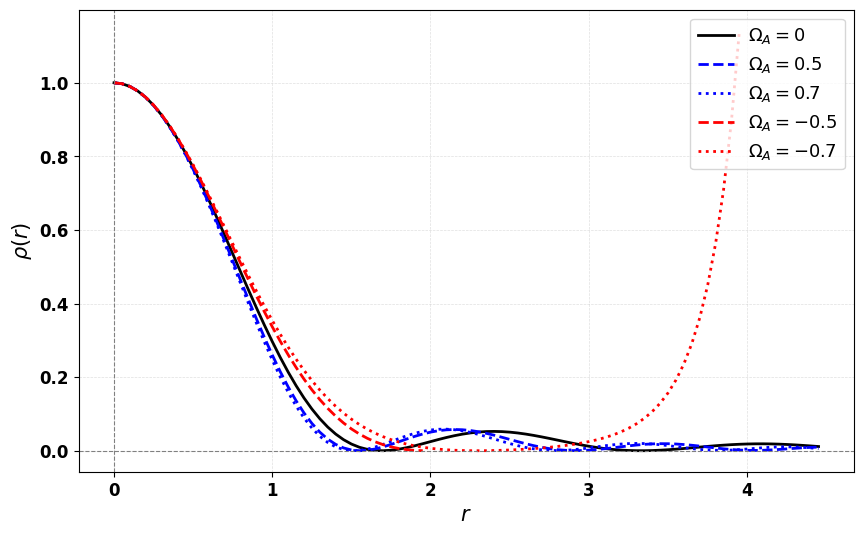}
        \subcaption{Interacting case ($\Omega_s=1$, $\Omega_q = 1$, $\Omega_p = 0$)}
        \label{fig:sub3}
    \end{minipage}%
    \hfill
    \begin{minipage}[t]{0.48\columnwidth}
        \centering
        \includegraphics[width=\columnwidth]{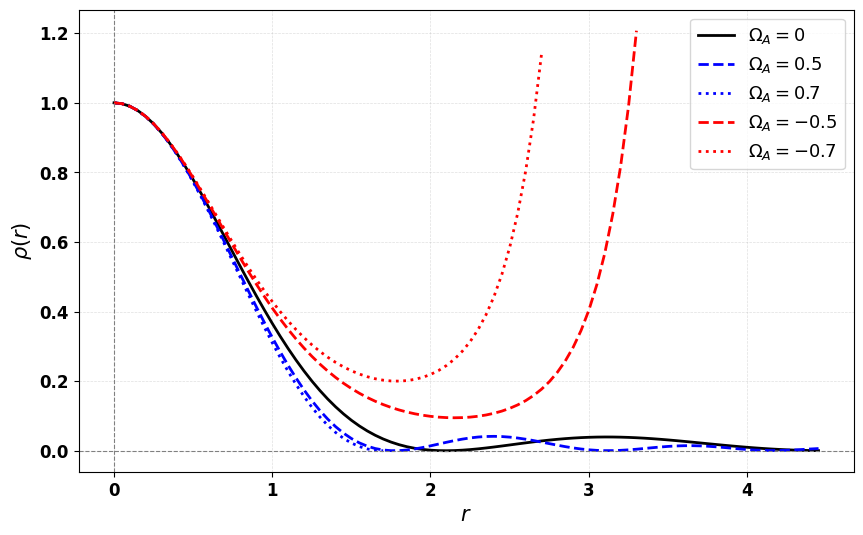}
        \subcaption{Interacting case ($\Omega_s=-1$, $\Omega_q = 1$, $\Omega_p = 0$)}
        \label{fig:sub4}
    \end{minipage}

    \caption{Radial density profiles for different parameter combinations in the pressureless case ($\Omega_P = 0$).}
    \label{fig:Q1P0}
\end{figure}


\subsection{Anisotropic Gross-Pitaevski-Poisson system}

This formulation represents the anisotropic extension of the system analyzed, for example, in \cite{chavanisI.84.043531} and can be derived from Eq. \eqref{general} by keeping $\Omega_s\neq0$ and imposing $\Omega_P=0$, thereby eliminating the pressure gradient contribution. Note, however, that this is formally equivalent to a polytropic configuration with $\gamma=2\,\,(n=1)$. 

Again, the specific case referred to as the "Quantum polytrope" corresponds to a subcase in which the boson-boson interaction is absent, i.e., $\Omega_s=0$. This restriction simplifies the system, reducing it to a scenario where equilibrium is maintained solely through quantum pressure and gravitational effects as done in the previous subsection. 

The possible freedom now within the free parameter space is to explore possible differences between attractive ($a_s<0$) or repulsive ($a_s>0$) interactions. At laboratory level, Feshbach resonance allows precise control of these interactions by tuning $a_s$ by applying an external field, such as a magnetic field. In this phenomenon, small variations in the external field tune the system close to a quasi-bound molecular state, causing large changes in the value of $a_s$, including allowing its sign to be reversed \cite{inouye1998observation,chin2010feshbach}.

In Fig. \ref{fig:Q1P0}, we activate the scattering length. 
The left panels of Fig. \ref{fig:Q1P0} illustrate cases where the repulsive boson-boson interaction is considered, with $\Omega_s=1$. We also explore negative scattering length scenarios in the right panels of this figure. $\Omega_s < 0$ values correspond to an attractive interaction between particles. In a BEC, this means that the bosons tend to cluster together rather than repel each other, which can lead to self-trapping or collapse.

In these scenarios, the interaction introduces additional structural effects that influence the equilibrium properties of the system. Specifically, when the repulsive boson-boson interaction is counterbalanced solely by the gravitational field, meaning $\Omega_s=1$ while $\Omega_q=\Omega_P=0$, stable, bound configurations cease to exist for values of $\Omega_A \gtrsim 0.1$. This behavior, depicted in the top left panel of Fig. \ref{fig:Q1P0}, highlights the threshold beyond which the anisotropic parameter disrupts equilibrium, leading to unstable configurations. These findings emphasize the intricate interplay between self-interactions and gravitational effects in shaping the stability of bosonic structures.

\begin{figure}[h]
    \centering
    \begin{minipage}[t]{0.48\columnwidth}
        \centering
        \includegraphics[width=\columnwidth]{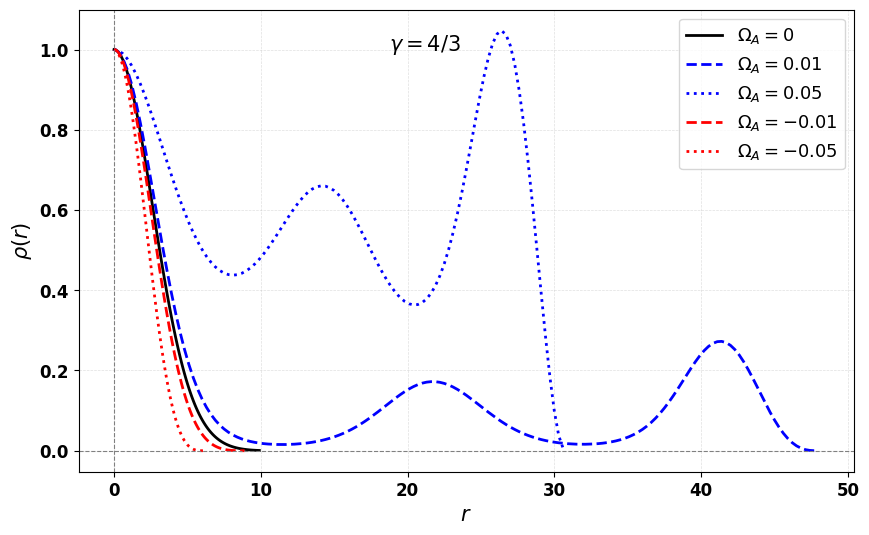}
    \end{minipage}%
    \hfill
    \begin{minipage}[t]{0.48\columnwidth}
        \centering
        \includegraphics[width=\columnwidth]{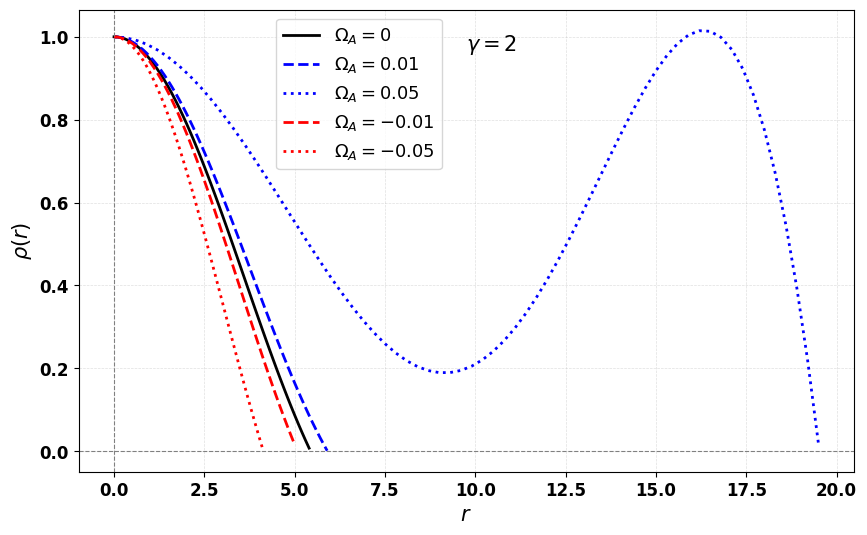}
    \end{minipage}
    \hfill
   \begin{minipage}[t]{0.48\columnwidth}
        \centering
        \includegraphics[width=\columnwidth]{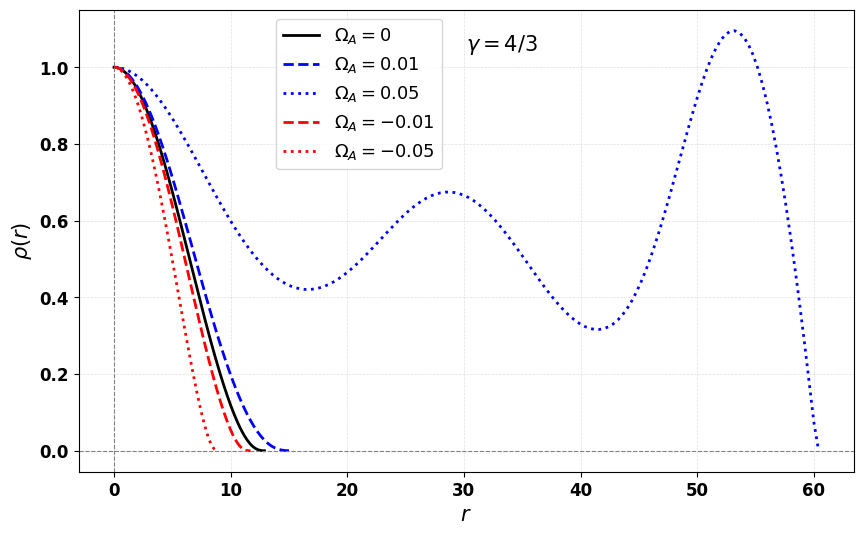}
    \end{minipage}%
    \hfill
    \begin{minipage}[t]{0.48\columnwidth}
        \centering
        \includegraphics[width=\columnwidth]{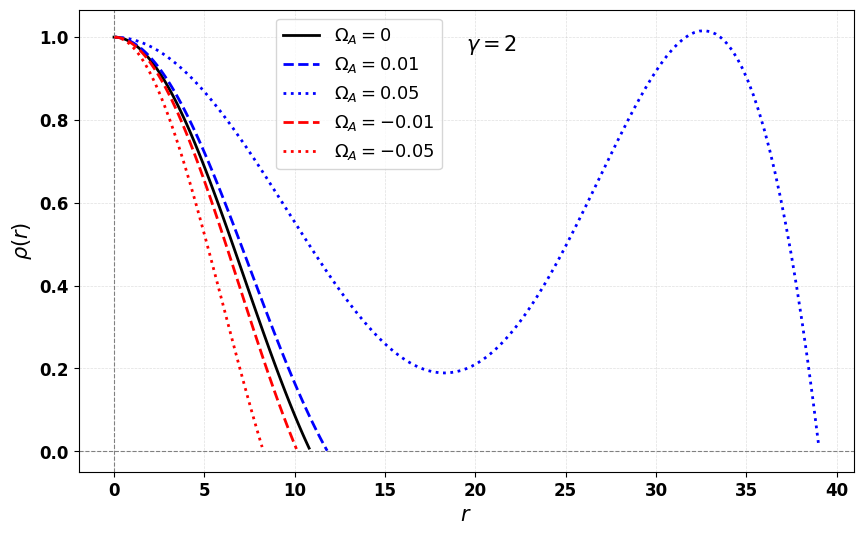}
    \end{minipage}
    \caption{Radial density profile in the interacting case with $\Omega_s = 1$ and $\Omega_q = 0$ (Thomas-Fermi Limit). Here $\Omega_p = 1$ adopting $\gamma=4/3$ (left panel) and $\gamma=2$ (right panel). Delta profile in the interacting case with $\Omega_s = 1$ and $\Omega_q = 0$ (Thomas-Fermi Limit).  When $\gamma=4/3$, the limiting situation occurs when $\Omega_A \gtrsim 2.0$. With $\gamma=2$, the function no longer reaches $\rho(r) = 0$ for values close to or less than $\Omega_A = 0.07$.}
    \label{fig:Omegas1Q0p1}
\end{figure}

In Fig. \ref{fig:Omegas1Q0p1}, the pressure gradient contribution is activated with different magnitudes, highlighting its effect on the system's equilibrium. In the top panels, the parameter is set to $\Omega_P=1$, representing a moderate pressure influence, while in the bottom panels, a significantly stronger contribution is considered, with $\Omega_P=10$. This increase in $\Omega_P$ leads to notable variations in the structural properties, influencing the stability and spatial distribution of the configuration. The figure illustrates how changes in the pressure gradient affect the overall behavior of the system. Increasing $\Omega_s$ leads to an increase in $R$, similar to the effect of raising the pressure term $\Omega_p$ in the classical case. However, the increase in $R$ is more subtle. The behavior previously observed when $\Omega_A = 0.01$ is not present here. Apparently, the limit occurs for $\gamma = 4/3$ when $\Omega_A > 1.0$, despite the unusual behavior at high values of $\Omega_A$, particularly for $\Omega_A > 0.09$. For $\gamma = 2$, the function does not reach zero density for values close to or exceeding $\Omega_A = 0.07$.

\section{Conclusions}

We have studied anisotropic hydrostatic equilibrium configurations under the assumption that, in addition to the outward pressure gradient and attractive gravitational effects, two additional quantum contributions—the quantum potential and the boson-boson interaction potential—contribute to the polytropic structure. The general equilibrium equation is presented in Eq. \eqref{general}, where each $\Omega_i\, (i=p,q,s,G,A)$ factor represents the magnitude of the aforementioned effects.

While the anisotropic factor $\Delta = P_t - P_r$ (defined according to \eqref{Delta}) is usually interpreted as an outward force contribution when positive ($\Delta > 0$), we identified the opposite effect in the presence of the quantum potential. Specifically, when $\Omega_q = \frac{\hslash^2}{2m^2} = 1$, a positive anisotropic factor $\Delta$ acts as an inward force, favoring clustering. This represents an exotic behavior not previously identified in the literature, to the best of our knowledge.

Furthermore, the inclusion of anisotropic features leads to an oscillatory radial density behavior that can be connected to similar features observed in simulated galactic rotation curves of ultralight dark matter halos \cite{Schive:2014dra,Schive:2014hza}, as also noted by \cite{Bar:2018acw}. We have therefore found a possible source of degeneracy between anisotropic pressure contributions and axion-like particle effects within clustered structures. We shall investigate this issue in more detail in future contributions.

\acknowledgments{The authors thank FAPEMIG, FAPES, CAPES and CNPq for financial support.}

\end{document}